\documentclass[runningheads]{llncs}

\usepackage{color}
\usepackage{graphicx}
\usepackage{amsmath}
\usepackage{amssymb}

\newcommand{\ovr}{\overline{r}}
\newcommand{\SA}{\mathit{SA}}
\newcommand{\LCP}{\mathit{LCP}}

\newcommand{\RMQ}{\mathit{RMQ}}
\newcommand{\PSV}{\mathit{PSV}}
\newcommand{\NSV}{\mathit{NSV}}

\title{Contextual Pattern Matching%
\thanks{Supported in part by Fondecyt grant 1200038 and 1260080, and Basal Funds FB0001, Chile.}}
\titlerunning{Contextual Pattern Matching}
\author{Gonzalo Navarro} %\orcidID{0000-0002-2286-741X}}
\authorrunning{G. Navarro}

\institute{CeBiB --- Center for Biotechnology and Bioengineering,\\
Department of Computer Science, University of Chile. \\
Beauchef 851, Santiago, Chile. \texttt{gnavarro@dcc.uchile.cl}}

\begin{document}

\maketitle

\begin{abstract}
The research on indexing repetitive string collections has focused on the same
search problems used for regular string collections, though they can make little
sense in this scenario. For example, the basic pattern matching query ``list 
all the positions where pattern $P$ appears'' can produce huge outputs when 
$P$ appears in an area shared by many documents. All those occurrences are 
essentially the same.

In this paper we propose a new query that can be more appropriate 
in these collections, which we call {\em contextual pattern matching}. The 
basic query of this type gives, in addition to $P$, a context length $\ell$, 
and asks to report the occurrences of all {\em distinct} strings $XPY$, with 
$|X|=|Y|=\ell$.

While this query is easily solved in optimal time and linear space, we focus
on using space related to the repetitiveness of the text collection and present
the first solution of this kind. Letting $\ovr$ be the maximum of the number of
runs in the BWT of the text $T[1..n]$ and of its reverse, our structure uses 
$O(\ovr\log(n/\ovr))$ space and finds the $c$ contextual 
occurrences $XPY$ of $(P,\ell)$ in time $O(|P| + c \log n)$.
We also show how, within space $O(\ovr)$, the problem can be solved 
in time $O((m+\lambda \cdot occ)\log\log n)$.
We give other space/time tradeoffs as well, for compressed and uncompressed
indexes.
\end{abstract}

\section{Introduction}

About a decade ago, it was realized that many of the fastest-growing text 
collections of the ``data deluge'' were highly repetitive \cite{MNSV09}. Since 
then, a number of 
research results have focused on developing indexes whose size is related to 
some good measure of compressibility for highly repetitive string collections 
\cite{Nav20}. Today one can find indexes built on measures like the size of
the Lempel-Ziv parse \cite{KN13,GGKNP14,FKP18,BEGV18}, of a grammar 
generating only the text \cite{CNspire12,TTS14}, of a string attractor 
\cite{NP18,CEKNP19}, the number of runs in the Burrows-Wheeler Transform 
(BWT) \cite{BW94} of the text \cite{MNSV09,GNP19}, or the size of an
automaton \cite{BBHMCE87} recognizing text substrings \cite{BCGPR15,BC17}.

All these indexes are devoted to the basic {\em pattern matching} query: given
a short pattern string $P[1..m]$, output all the $occ$ positions where it 
occurs in the text $T[1..n]$. Some indexes have managed to solve this problem 
in optimal time, $O(m+occ)$, using space bounded by some function of the above 
measures \cite{BC17,GNP19}, whereas others have low polylogarithmic factors
multiplying $m$ or $occ$. 

While very reasonable in general, this query can be pretty useless in a highly
repetitive text collection. A pattern $P$ that appears inside a highly repeated
text area will be reported myriad times, wasting a lot of effort to produce and
to handle the result. We are not aware of many efforts to propose queries that 
are better adapted to a scenario of high repetitiveness.

In this paper we make a first step in this direction. We propose a query called
{\em contextual pattern matching} which, in addition to $P$, gives a context
length $\ell$. We then want one element of output per distinct context where
$P$ appears, that is, all the positions where $P$ appears preceded by the same 
string $X$ of length $\ell$ and followed by the same string $Y$ of length 
$\ell$ shall be reported only once. 

\begin{definition}
The {\em contextual pattern matching} problem on a text $T[1..n]$ is, given a 
pair $(P[1..m],\ell)$, return a position in $T$ for each of the $c$ distinct
strings $XPY$ occurring in $T$, for all $X,Y$ such that $|X|=|Y|=\ell$. 
For the occurrences near the extremes of $T$, assume $T$ is preceded and 
followed by $\ell$ copies of the special symbol $\$$, which cannot appear in
$P$.
\end{definition}

It is not hard to solve this query in optimal time 
$O(m+c)$ if we use linear space, $O(n)$, by using suffix trees \cite{Wei73}
and other linear-space auxiliary structures.
% el mapeo se me complica con los implicit nodes
%for $T$ and for its reverse $T^{rev}$, with pointers mapping their explicit 
%nodes to the corresponding (implicit or explicit) node in the other tree. At 
%query time, find the suffix tree locus $v$ of $P$ in time $O(m)$ and traverse 
%its subtree to find its $k \le c$ (implicit or explicit) descendant nodes of 
%string depth $m+\ell$, in time $O(k)$. Map each such node $u$ to the
%suffix tree of $T^{rev}$ and collect its descendant (implicit or explicit)
%nodes at string depth $m+2\ell$. Map back each such node $w$ to the suffix
%tree of $T$, reporting the corresponding suffix array range.
We are interested, however, in using space related to a relevant repetitiveness
measure. We show that, if we call $\ovr$ the maximum of the number of 
equal-letter runs in the BWT of $T$ or its reverse, then a data structure
using $O(\ovr\log(n/\ovr))$ space can solve contextual pattern matching in time
$O(m + c \log n)$; and within space $O(\ovr)$ we do it
in time $O((m+\lambda \cdot occ)\log\log n)$.
We also show how any compressed text index can be 
extended with $O(n)$ {\em bits} and efficiently solve this query; this can be
interesting for mildly repetitive texts.

\section{Preliminaries}
\label{sec:prelim}

We index a text $T[0..n]$ over alphabet $[1..\sigma]$, where $T[0]=T[n]=\$$ 
is a special terminator smaller than all the other alphabet symbols. The {\em
suffix array} \cite{MM93} $\SA[1..n]$ of $T$ lists all the suffixes $T[i..n]$ for $i\ge 1$ in
lexicographic order, and the {\em LCP array}, $\LCP[1..n]$, gives the length of 
the longest common prefix between consecutive suffix array entries, 
$\LCP[i] = lcp(T[\SA[i]..n],T[\SA[i-1]..n])$. 
%We use the RAM model of computation with $w$-bit words, assuming $w=\Omega(\log n)$.

One relevant measure of repetitiveness is called $r$, the number of equal-letter
runs in the Burrows-Wheeler Transform (BWT) of $T[1..n]$. The BWT \cite{BW94} is
a reordering of the symbols of $T$ obtained by collecting the symbol preceding 
the lexicographically sorted suffixes of $T$. That is, if $\SA[1..n]$ is the 
suffix array of $T$, then $BWT[i] = T[\SA[i]-1]$. For example, it is known that
$r = O(\gamma \log^2 n)$ \cite{KK19}, where $\gamma$ is the smallest attractor
of $T$ \cite{KP18}.

Gagie et al.~\cite[Sec.~5.2--5.4 \& 6.3]{GNP19} describe data structures of 
size $O(r\log(n/r))$ that can find the suffix array range of any pattern 
$P[1..m]$ in time $O(m)$, and that can compute any entry $\SA[i]$, 
$\SA^{-1}[i]$, and $\LCP[i]$, in time $O(\log(n/r))$. Further, they can compute
the following queries on $\LCP$:\footnote{Our $\PSV$ and $\NSV$ are called
$\PSV'$ and $\NSV'$ in there.}
\begin{itemize}
\item $\RMQ(i,j) = \textrm{argmin}_{i\le k\le j} \LCP[k]$ in time
$O(\log(n/r))$.
\item $\PSV(p,d) = \max(\{q < p, \LCP[q] < d\} \cup \{0\})$, in time
$O(\log(n/r)+\log\log_w r)$.
\item $\NSV(p,d) = \min(\{q > p, \LCP[q] < d\} \cup \{n + 1\})$, in time
$O(\log(n/r)+\log\log_w r)$.
\end{itemize}
%The $O(\log(n/r))$-space data structures are binary 
%context-free grammars of height $O(\log(n/r))$ built on the differential 
%versions of the arrays, for example,
%$\DSA[i]=\SA[i]-\SA[i-1]$ in the case of the suffix array. The grammars exploit
%the
%fact that these differential sequences inherit the repetitiveness of the text.

\section{Our Solution}

We present a suffix-array-oriented solution that solves a stronger variant of
the problem: we give the $c$ suffix array ranges of all the distinct contexts
$XPY$ where $P$ occurs in $T$. We can then report one text position for
each, but also determine how many times each context occurs, and report its
occurrences one by one.

We store the described data structures of Gagie et al.~\cite{GNP19} for
both $T[0..n]$ and its reverse $T^{rev}[0..n]$. We call $r$ and $r'$ the number
of equal-letter runs in the BWT of $T$ and of $T^{rev}$, respectively, and 
$\ovr=\max(r,r')$. Therefore the structures we use take space 
$O(\ovr\log(n/\ovr))$.
The general strategy to solve a query $(P[1..m],\ell)$ is as follows:

\begin{enumerate}
\item We first find, in $O(m)$ time, the suffix array range
$[rs..re]$ of $P^{rev}$ (i.e., $P$ read backwards) in the suffix array $\SA'$ of $T^{rev}$.
\item We then partition $[rs..re]$ into $k \le c$ maximal consecutive intervals 
$[rs_i,re_i]$ where the suffixes in each interval share their first $m+\ell$ 
symbols, that is, $T^{rev}[\SA'[p]..\SA'[p]+m+\ell-1] = P^{rev}X_i^{rev}$ for all
$rs_i \le p \le re_i$.
\item We map each interval $\SA'[rs_i,re_i]$ to the interval $\SA[ds_i..de_i]$ 
corresponding to the suffixes that start with $X_i P$.
\item We partition each interval $\SA[ds_i..de_i]$ into $k_i$ maximal consecutive
subintervals $\SA[ds_i^j..de_i^j]$ where the suffixes in each subinterval share
their first $m+2\ell$ symbols, $T[\SA[p]..\SA[p]+m+2\ell-1] = X_i P Y_j$ for all
$ds_i^j \le p \le de_i^j$.
\item We report the $c = \sum_{i=1}^k k_i$ resulting subintervals 
$\SA[ds_i^j..de_i^j]$ and, if desired, a text position $\SA[p]$ with
$ds_i^j \le p \le de_i^j$ for each.

\end{enumerate}

We now solve the two nonobvious subproblems of our general strategy. The first,
in points 2 and 4, is to partition a suffix array interval into subintervals
of suffixes sharing their first $t$ symbols. The second, in point 3, is how to
map an interval of the suffix array of $T^{rev}$ into the corresponding interval
in the suffix array of $T$. The solutions we find have a complexity of
$O(\log n)$ per item output, which leads to our promised result.

\begin{theorem}
Let $T$ be a text of length $n$, and let $\ovr$ be the maximum of the number of
equal letter runs of its BWT and the BWT of its reverse. Then there is a data
structure of size $O(\ovr \log(n/\ovr))$ that finds the $c$ contextual 
occurrences of $(P[1..m],\ell)$ in time $O(m + c \log n)$.
\end{theorem}

The data structures \cite{GNP19} can be built in $O(n)$ time and space, or in
$O(n\log n)$ time and $O(\ovr\log(n/\ovr))$ space, the same as the final space
of the structures. 

\paragraph{Example.}

Figure~\ref{figure:ex} shows an example on the text $T[0..17]=\mathsf{\$alabaralalabarda\$}$, where we search for $P=\mathsf{a}$ with context length $\ell=1$.
Step 1 finds the interval $\SA'[rs..re] = \SA'[2..9]$ of all the occurrences of $P^{rev}=\mathsf{a}$ on $T^{rev}$. Step 2 finds the places where 
$\LCP'[p]<m+\ell=2$ (see Section~\ref{sec:sub1}), for $p \in [2..9]$, namely $2, 3, 5, 6, 9$. These are the starting positions of the intervals $[rs_i,re_i] = [2,2], [3,4], [5,5], [6,8],
[9,9]$, and correspond to the contexts $P^{rev} X_i^{rev} = \mathsf{a\$}, \mathsf{ab}, \mathsf{ad},$ $\mathsf{al}, \mathsf{ar}$. 
Step 3 maps those intervals to $\SA$ (see Section~\ref{sec:sub2}), $[ds_i,de_i] = [5,5], [10,11],$ $[12,12], [13,15], [16,16]$; they retain the same order of $\SA'$ only because $\ell=1$.
Step 4 splits each interval at subintervals starting wherever $\LCP[p] < m+2\ell = 3$, namely positions $5, 10,12,13,15,16$. Therefore, the resulting
subintervals (i.e., the output) are $[5,5]$, $[10,11]$, $[12,12]$, $[13,14]$, $[15,15]$, $[16,16]$, corresponding to the contexts \textsf{\$al},
\textsf{bar}, \textsf{da\$}, \textsf{lab}, \textsf{lal}, \textsf{ral}.

We also show the array $C$ used in Section~\ref{sec:larger}; note that each $ds_i$ corresponds to mapping the minimum position of $C$
in $[rs_i,re_i]$.

\bigskip

\begin{figure}[tb]
\begin{center}
\begin{tabular}{|@{~}r@{~}|@{~}c@{~}|@{~}c@{~}|@{~}c@{~}|@{~}c@{~}|@{~}c@{~}|@{~}c@{~}|@{~}c@{~}|@{~}c@{~}|@{~}c@{~}|@{~}c@{~}|@{~}c@{~}|@{~}c@{~}|@{~}c@{~}|@{~}c@{~}|@{~}c@{~}|@{~}c@{~}|@{~}c@{~}|@{~}c@{~}|}
\hline
$n$          & 0 & 1 & 2 & 3 & 4 & 5 & 6 & 7 & 8 & 9 & 10& 11& 12& 13& 14& 15& 16& 17 \\
\hline
$T$          & \textsf{\$} & \textsf{a} & \textsf{l} & \textsf{a} & \textsf{b} & \textsf{a} & \textsf{r} & \textsf{a} & \textsf{l} & \textsf{a} & \textsf{l} & \textsf{a} & \textsf{b} & \textsf{a} & \textsf{r} & \textsf{d} & \textsf{a} & \textsf{\$} \\
\hline
$\SA$          &   & 17& 16& 3 & 11& 1 & 9 & 7 & 5 & 13& 4 & 12& 15& 2 & 10& 8 & 6 & 14 \\
\hline
             \multicolumn{6}{}{}  & \multicolumn{1}{c}{$\!\!\!\underbrace{\hspace{6mm}}^{ds_1 \hfill de_1}$} & \multicolumn{4}{c}{} &
				    \multicolumn{2}{c}{$\!\!\!\underbrace{\hspace{10mm}}^{ds_2 \hfill de_2}$} &
				    \multicolumn{1}{c}{$\!\!\!\!\underbrace{\hspace{6mm}}^{ds_3 \hfill de_3}$} &
				    \multicolumn{3}{c}{$\!\!\!\!\underbrace{\hspace{16mm}}^{ds_4 \hfill de_4}$} &
				    \multicolumn{1}{c}{$\!\!\!\underbrace{\hspace{6mm}}^{ds_5 \hfill de_5}$} \\
             \multicolumn{6}{}{}  & \multicolumn{1}{c}{$\!\!\!\underbrace{\hspace{6mm}}^{\mathsf{\$al}}$} & \multicolumn{4}{c}{} &
				    \multicolumn{2}{c}{$\!\!\!\underbrace{\hspace{10mm}}^{\mathsf{bar}}$} &
				    \multicolumn{1}{c}{$\!\!\!\!\underbrace{\hspace{6mm}}^{\mathsf{da\$}}$} &
				    \multicolumn{2}{c}{$\!\!\!\!\underbrace{\hspace{10mm}}^{\mathsf{lab}}$} &
				    \multicolumn{1}{c}{$\!\!\!\!\underbrace{\hspace{6mm}}^{\mathsf{lal}}$} &
				    \multicolumn{1}{c}{$\!\!\!\underbrace{\hspace{6mm}}^{\mathsf{ral}}$} \\
\hline
$\LCP$	     &   & 0 & 0 & 1 & 4 & 1 & 6 & 3 & 1 & 2 & 0 & 3 & 0 & 0 & 5 & 2 & 0 & 1  \\
\hline
             \multicolumn{6}{}{} & \multicolumn{1}{c}{$\uparrow_\mathsf{\$al}$} & \multicolumn{4}{c}{} &
				   \multicolumn{1}{c}{$\uparrow_\mathsf{bar}$} & \multicolumn{1}{c}{} &
				   \multicolumn{1}{c}{$\uparrow_\mathsf{da\$}$} &
                                   \multicolumn{1}{c}{$\uparrow_\mathsf{lab}$} & \multicolumn{1}{c}{} &
				   \multicolumn{1}{c}{$\uparrow_\mathsf{lal}$} & 
				   \multicolumn{1}{c}{$\uparrow_\mathsf{ral}$}  \\
\hline
$T^{rev}$    & \textsf{\$} & \textsf{a} & \textsf{d} & \textsf{r} & \textsf{a} & \textsf{b} & \textsf{a} & \textsf{l} & \textsf{a} & \textsf{l} & \textsf{a} & \textsf{r} & \textsf{a} & \textsf{b} & \textsf{a} & \textsf{l} & \textsf{a} & \textsf{\$ } \\
\hline
$\SA'$         &   & 17& 16& 12& 4 & 1 & 14& 6 & 8 & 10& 13& 5 & 2 & 15& 7 & 9 & 11& 3  \\
\hline
             \multicolumn{3}{}{}  & \multicolumn{8}{c}{$\!\!\!\underbrace{\hspace{4.7cm}}^{rs \hfill re}$} \\
             \multicolumn{3}{}{}  & \multicolumn{1}{c}{$\!\!\!\underbrace{\hspace{6mm}}^{rs_1 \hfill re_1}$} &
				    \multicolumn{2}{c}{$\!\!\!\underbrace{\hspace{10mm}}^{rs_2 \hfill re_2}$} &
				    \multicolumn{1}{c}{$\!\!\!\!\underbrace{\hspace{6mm}}^{rs_3 \hfill re_3}$} &
				    \multicolumn{3}{c}{$\!\!\!\!\underbrace{\hspace{12mm}}^{rs_4 \hfill re_4}$} &
				    \multicolumn{1}{c}{$\!\!\!\underbrace{\hspace{6mm}}^{rs_5 \hfill re_5}$} \\
\hline
$\LCP'$       &   & 0 & 0 & 1 & 5 & 1 & 1 & 3 & 3 & 1 & 0 & 4 & 0 & 0 & 2 & 2 & 0 & 6  \\
\hline
             \multicolumn{3}{}{} & \multicolumn{1}{c}{$\uparrow_\mathsf{a\$}$} &
				   \multicolumn{1}{c}{$\uparrow_\mathsf{ab}$} & \multicolumn{1}{c}{} &
				   \multicolumn{1}{c}{$\uparrow_\mathsf{ad}$} &
                                   \multicolumn{1}{c}{$\uparrow_\mathsf{al}$} & \multicolumn{2}{c}{} &
				   \multicolumn{1}{c}{$\uparrow_\mathsf{ar}$}  \\
\hline
$C$          &   & --& 5 & 8 & 9 & 2 & 3 & 4 & 6 & 7 & 10& 11& 12& 13& 14& 15& 16& 17 \\
\hline
\end{tabular}
\end{center}
\caption{Example trace.}
\label{figure:ex}
\end{figure}

\subsection{Partitioning a suffix array interval} \label{sec:sub1}

Given a range $[s..e]$ of the suffix array of a string $S$, and a length $t$, 
we must partition it into maximal subranges $[s_1..e_1],\ldots,
[s_k..e_k]$ where the suffixes starting in each subrange share their first $t$ 
symbols. 

Note that $s_1=s$ and $s_2,\ldots,s_k$ are the values in $[s..e]$ where
$\LCP[i] < t$, where $\LCP$ is the LCP array of $S$. We use the standard 
procedure for 3-sided queries to find all those positions $s_i$: compute 
$p = \RMQ(s,e)$ and, if $\LCP[p] < t$, recurse on $[s,p-1]$, report $p$, 
and recurse on $[p+1,e]$. This process requires $O(k)$ computations of 
$\RMQ$ and accesses to $\LCP$ to find $s_2,\ldots,s_k$ in order, each of 
which takes time $O(\log(n/r))$, where $r$ is the number of runs in the BWT
of $S$. 

Since we use this technique for $S=T$ and $S=T^{rev}$, the total cost is 
$O(k \max(\log(n/r),\log(n/r'))) \subseteq O(k\log n)$.

\subsection{Mapping suffix array intervals} \label{sec:sub2}

Given the suffix array interval $\SA'[s'..e']$ of $T^{rev}$, consisting of all 
the suffixes that start with a string of length $t$, we want to find the
corresponding suffix array interval $\SA[s..e]$ of $T$. With the suffix
array $\SA'$ of $T^{rev}$ and the inverse suffix array $\SA^{-1}$ of $T$, we can
translate any such suffix, say $p = \SA^{-1}[n-\SA'[s']-(m+\ell-1)]$ (or
$p=\SA^{-1}[1]$ if $n-\SA'[s']-(m+\ell-1) \le 0$). We indeed store the 
structures to compute those in time $O(\log(n/r)+\log(n/r')) \subseteq
O(\log n)$ (Section~\ref{sec:prelim}).

We know that $s \le p \le e$, so the task is to extend $p$ in both directions:
$s \le p$ is the largest position where $\LCP[s] < t$ and $e \ge p$ is the
smallest position where $\LCP[e+1] < t$. Those are, precisely, the operations
$\PSV(p,t)$ and $\NSV(p,t)-1$ that our structures on $\LCP$ compute in time 
$O(\log(n/r)+\log\log_w r) \subseteq O(\log n)$ (Section~\ref{sec:prelim}
again).

\subsection{Running on Other Indexes} \label{sec:larger}

If we are willing to store uncompressed data structures of $O(n)$ space, 
we can find the interval of point (1) in RAM-optimal time $O(m/\log_\sigma n)$ 
using an enhanced suffix tree \cite{NN17} on
$T^{rev}$. The $k$ intervals $[rs_i,re_i]$ of point (2) can be found in 
$O(k)$ time using range minimum queries $\RMQ'(i,j)$ on the LCP array of 
$T^{rev}$, $\LCP'$. Such queries take constant time using $2n+o(n)$ bits of 
space \cite{FH11}. Each such interval $\SA'[rs_i,re_i]$ can then be mapped 
(point 3) to $\SA[ds_i,de_i]$ by storing an array $C[1..n]$ with 
$C[i] = \SA^{-1}[n-\SA'[i]]$ and building an $\RMQ_C$ data structure on $C$, 
so that $ds_i = \SA^{-1}[n-\SA'[\RMQ_C(rs_i,re_i)]-(m+\ell-1)]$ and 
$de_i = ds_i + (re_i-rs_i)$. (Note that we build $C$ on the values 
$\SA^{-1}[n-\SA'[i]]$, not $\SA^{-1}[n-\SA'[i]-(m-\ell+1)]$, because
the latter depend on $\ell$ and all the suffixes in this range share their 
first $m+\ell$ symbols anyway, so the lexicographic comparison is the same.)
Finally, point (4) on each $\SA[ds_i,ds_i]$ is solved as for point (2), now on 
the constant-time $\RMQ$ structure for the LCP array of $T$. 
The total time is then the optimal $O(m/\log_\sigma n+c)$.

\begin{theorem}
Let $T$ be a text of length $n$ over an alphabet of size $\sigma$. Then there 
is a data of size $O(n)$ that finds the $c$ contextual occurrences of 
$(P[1..m],\ell)$ in time $O(m/\log_\sigma n+c)$.
\end{theorem}

More generally, if we have an index that finds the suffix array range $[rs..re]$
for $P$ in $T^{rev}$, and can extract any cell of $\SA$, $\SA^{-1}$, and $\SA'$,
we can use it for contextual reporting using our general solution.
We need $O(n)$ extra bits for the various $\RMQ$ data structures. Note
we do not need to store $C$ explicitly because we can simulate it using $\SA'$
and $\SA^{-1}$. Further, the arrays $\LCP'$ and $\LCP$ are simulated with other 
$2n+o(n)$ bits if we have access to $\SA'$ and $\SA$ \cite{Sad07}.
We then have the following result.
% 6n bits in total para rmq en LCP',LCP,C y 4n para los LCP' y LCP mismos
% 10n en total

\begin{theorem}
Let $T$ be a text of length $n$ and an index on $T^{rev}$ using $\cal S$ bits 
of space that finds the suffix array range of $P[1..m]$ in time $t_s(m)$, and 
computes any cell of $\SA$, $\SA'$, or $\SA^{-1}$ in time $t_\SA$, where $\SA$ and $\SA'$ 
are the suffix arrays of $T$ and $T^{rev}$, respectively. Then there is a data
structure using ${\cal S}+O(n)$ bits of space that finds the $c$ contextual 
occurrences of $(P[1..m],\ell)$ in time $O(t_s(m)+c\,t_\SA)$.
\end{theorem}

Building on an index \cite{BN13} that uses $nH_k(T^{rev})+o(n\log\sigma)+O(n)$ 
bits of space for any $k < \alpha\log_\sigma n$ and constant $0<\alpha<1$, 
where $H_k(S) < \log\sigma$ 
is the $k$th order empirical entropy of string $S$ \cite{Man01},
we have $t_s(m)=O(m)$ and $t_\SA = O(\log n)$. The index provides access to 
$\SA'$ and $(\SA')^{-1}$ by storing their values at regular intervals of $T^{rev}$,
of length $s=\Theta(\log n)$ in our case, and marking the sampled positions of 
$\SA'$ in a bitvector. It provides a way to move in constant time from $i$ such 
that $\SA'[i]=j$ to $i'=LF(i)$ such that $\SA'[i']=j-1$. Thus, if $\SA'[i]$ is not 
sampled, it can move $s' < s$ times until finding a sampled cell 
$\SA'[LF^{s'}(i)]=j'$, and then $\SA'[i]=j'+s'$. The same $LF$ function is used 
$j'-j < s$ times, for $j'=\lceil j/s \rceil\cdot s$, to find $(\SA')^{-1}[j]$, 
by starting from the sampled value $(\SA')^{-1}[j']$ and tracing it back to 
$(\SA')^{-1}[j] = LF^{j'-j}((\SA')^{-1}[j'])$. Enhancing it to computing values of
$\SA$ and $\SA^{-1}$ (which correspond to $T$) requires to store their 
sampled values as well, because $T^{rev}[j] = T[n-j]$. Finally, because 
$H_k(T) = H_k(T^{rev})$ \cite[Sec.~11.3.2]{Nav16}, we have the following result.

\begin{theorem}
Let $T$ be a text of length $n$ over an alphabet of size $\sigma$, with $k$th 
order empirical entropy $H_k(T)$, for any $k < \alpha\log_\sigma n$ and 
constant $0<\alpha<1$. Then there is a data structure of 
$nH_k(T)+o(n\log\sigma)+O(n)$ bits that finds the $c$ contextual 
occurrences of $(P[1..m],\ell)$ in time $O(m + c \log n)$.
\end{theorem}

We can speed up this index by using {\em compact} space, $O(n\log\sigma)$ bits
(i.e., proportional to a plain representation of $T$). In this case, any cell
of $\SA$ or $\SA^{-1}$ (and of $\SA'$ by building the structures on $T^{rev}$ as well)
can be computed in time $O(\log_\sigma ^\epsilon n)$ for any constant 
$\epsilon>0$ \cite{GV06}. Further, this index finds the suffix array interval
of $P$ in almost RAM-optimal time, $O(m/\log_\sigma n + \log_\sigma^\epsilon n)$.

\begin{theorem}
Let $T$ be a text of length $n$ over an alphabet of size $\sigma$.
Then there is a data structure using $O(n\log\sigma)$ bits that finds
the $c$ contextual occurrences of $(P[1..m],\ell)$ in time 
$O(m/\log_\sigma n + (c+1) \log_\sigma^\epsilon n)$, for any constant 
$\epsilon > 0$.
\end{theorem}

\section{A Solution within $O(\ovr)$ Space}

We now make use of run-length bidirectional FM-indexes
\cite{BC19,BC20} to obtain a more compressed index which, in exchange, 
pays time proportional to $\lambda$ per occurrence (and outputs
the context string, which cannot be done in less time).

A bidirectional FM-index \cite{bidir} is, in principle, composed of the BWT 
$L$ of $T$ and 
the BWT $L'$ of $T^{rev}$. Those BWTs have $r$ and $r'$ equal-letter runs, 
respectively. We represent them as follows \cite{MN05,MNSV08} (we describe a 
representation of $L$ using $O(r)$ space; the one of $L'$ using $O(r')$ 
space is analogous).

\begin{enumerate}
\item The array $S[1..r]$ concatenating one copy of the letter of each run.
We add a data structure \cite{GMR06} that computes $rank_c(S,i)$ in time 
$O(\log\log\sigma)$, which is the number of times $c$ occurs in $S[1..i]$. 
\item The set $B$ of the $r$ positions where the runs start in $L$. This
set is represented using a predecessor data structure that uses $O(r)$ space and
answers queries in time $O(\log\log n)$; note $n$ is the universe size. Each 
element of $B$ also stores its rank in $B$ (i.e., $j$ if it is the $j$th run).
\item The array $D[1..\sigma]$ where $D[c]$ tells how many runs of 
symbols less than $c$ are there in $L$.
\item The array $B'[1..r]$ of the increasing positions where the runs of $L$
start after we sort them increasingly and stably by their letter.
\end{enumerate}

The key operation to search for $P$ using the FM-index is $LF(c,i)$. This is 
defined as follows \cite{FM00,FM05}:
\[ LF(c,i) ~=~ C[c] + rank_c(L,i), \]
where $C[c]$ counts how many symbols less than $c$ are there in $L$.
With our representation, $LF(c,i)$ can be computed in $O(\log\log n)$ time, as 
follows \cite{MN05}:
\begin{enumerate}
\item Compute $i' = pred(i)$ on $B$ to determine where the run of $i$ starts.
\item Letting $j$ be the rank of $i'$ in $B$, compute $k = rank_c(S,j)$.
\item If $c \neq S[j]$, then $LF(c,i) = B'[D[c]+k+1]-1$; otherwise 
$L(c,i) = B'[D[c]+k] + i-i'$.
\end{enumerate}

Searching for $P[1..m]$ using the BWT proceeds backwards: we start with the 
range $\SA[1..n]$ of the empty suffix of $P$ and, given the range $\SA[sp..ep]$
of $P[i+1..m]$ obtain the range of $P[i..m]$ with the following {\em backward 
step}: $sp \gets LF(P[i],sp-1)+1$ and $ep \gets LF(P[i],ep)$. We can then 
obtain the suffix array range of $P$ (or, equivalently, the range in $L$)
in time $O(m\log\log n)$.

We then start by obtaining the ranges $L[sp..ep]$ and $L'[sp'..ep']$ of $P$
in the BWTs of $T$ and $T^{rev}$, respectively. To obtain the contextual 
occurrences of $P$ in $T$, we will extend the ranges leftwards and rightwards, 
by all the possible sequences of $\lambda$ symbols. To do so, we need to
maintain all the time the range of every (partial) context $XPY$ in both $L$ and
$L'$. Bi-directional FM-indices achieve this as follows \cite{bidir}: when we
extend $P$ to the left by $c$, we use a backward step to find the new interval
$L[sp_c,ep_c]$ of $cP$. Then, the new interval of $P^{rev}c$ in $L'$ is 
$L'[sp'+s+1..sp'+s+o]$, where
$s$ is the number of symbols less than $c$ in $L[sp..ep]$ and $o = ep_c-sp_c+1$.
The value $s$ can be computed in time $O(\sigma)$ on small alphabets 
symbol by symbol, or in $O(\log\sigma)$ time on a wavelet tree \cite{GGV03,Nav13}
representation
of $L$. On run-length representations, however, this is not so simple 
\cite{BC19,BC20,ANS22}. Because we have to extend $P$ by every possible symbol,
however, the problem of computing $s$ vanishes. All we need is to extract the 
distinct symbols in $L[sp..ep]$ in time proportional to their number, and in
increasing order.

To do this with our representation, we map $L[sp..ep]$ to $S[i..j]$, where $i$
and $j$ are the ranks of $pred(sp)$ and $pred(ep)$, respectively. Thus the 
distinct symbols in $L[sp..ep]$ are the distinct symbols in $S[i..j]$. We then
use Muthukrishnan's optimal document listing algorithm \cite{Mut02} to obtain
the distinct values in $S[i..j]$, in constant time per value retrieved (this
algorithm requires that we precompute some $O(\sigma) \subseteq O(r)$ space 
data structures on top of $S$). We can insert the distinct symbols (which are
output in an arbitrary order) into a van Emde Boas tree with universe size 
(and hence space) $O(\sigma)$, and retrieve them in order with successor 
queries, in time $O(\log\log\sigma)$ per symbol retrieved.

We now have the distinct symbols $c_1 < c_2 < \cdots < c_k$ that precede the
current context in $T$. We start with $s \gets 0$ and compute $L[sp_1..ep_1]$
with a backward step for $c_1$ from $L[sp..ep]$. The corresponding range in
the BWT of $T^{rev}$ is $L'[sp'+s..sp'+s+(ep_1-sp_1)]$. Then we increase
$s \gets s + (ep_1-sp_1+1)$ and compute $L[sp_2..ep_2]$ with a backward step
for $c_2$ from $L[sp..ep]$. The corresponding range in $L'$ is 
$L'[sp'+s..sp'+s+(ep_2-sp_2)]$, and so on. This explores all the left-contexts
of length $1$. If $c_1 = \$$, this means we are in the beginning of $T$ and
thus do not extend that context anymore; otherwise we continue until obtaining
all the left-contexts up to length $\lambda$. For each such context $XP$, we now
proceed analogously extending the context to the right. Now the interval of 
$L'$ is updated using backward steps and that of $L$ is restricted using $s$.
At the end, we have spent $O(\lambda \log\log n)$ time to find each distinct
context
$XPY$. We can output the explicit context, and also report its (suffix array)
range in $L$.

\begin{theorem}
Let $T$ be a text of length $n$, and let $\ovr$ be the maximum of the number of
equal letter runs of its BWT and the BWT of its reverse. Then there is a data
structure of size $O(\ovr)$ that finds the $c$ contextual
occurrences of $(P[1..m],\ell)$, outputting their content, in time 
$O((m + \lambda \cdot c) \log\log n)$.
\end{theorem}

\section{Conclusions}

We have proposed a query that should be more meaningful than standard pattern
locating in the case of highly repetitive text collections. Instead of simply
locating all the positions of $T[1..n]$ where $P[1..m]$ appears, we give a 
context length $\ell$ and ask for the occurrences of all the $c$ distinct 
strings 
$XPY$ in the text, for any $X,Y$ where $|X|=|Y|=\ell$. If $P$ occurs inside a 
highly repeated substring, many essentially identical occurrences will be 
reported one by one with the standard locating, whereas we will report only a 
single suffix array range comprising all the occurrences of the same context 
$XPY$.

While the query can be solved in $O(n)$ space and RAM-optimal 
$O(m/\log_\sigma n +c)$ time, we focus on using space proportional to
the repetitiveness of $T$. We use one such measure, the number $r(S)$ of 
equal-letter runs of the Burrows-Wheeler Transform of the string $S$. Within
space $O(\ovr\log(n/\ovr))$, where $\ovr = \max(r(T),r(T^{rev}))$, we solve
the problem in time $O(m + occ \log n)$, and within space $O(\ovr)$ we do it 
in time $O((m+\lambda \cdot occ)\log\log n)$. We also show how to adapt
our general strategy to any compressed text index.

This is a first step towards studying queries that make more sense on highly
repetitive text collections, possibly deviating from the classical ones used
for regular collections.
Some relevant remaining questions are: Can the obtained space/time tradeoffs be 
improved? Are there other relevant and challenging queries that are better
suited to highly repetitive text collections?

\bibliographystyle{splncs04}
\bibliography{paper}

\end{document}